# Evolutionary Approach to S-box Generation: Optimizing Nonlinear Substitutions in Symmetric Ciphers

Oleksandr Kuznetsov [1,2], Nikolay Poluyanenko [2], Emanuele Frontoni [3], Marco Arnesano [1], and Oleksii Smirnov [4]

[1] *Faculty of Engineering, eCampus University, Via Isimbardi 10, Novedrate (CO), 22060, Italy*
[2] *V. N. Karazin Kharkiv National University, 4 Svobody Sq., 61022 Kharkiv, Ukraine*
[3] *Department of Political Sciences, Communication and International Relations, University of Macerata, Via Crescimbeni, 30/32, 62100 Macerata, Italy*
[4] *Department of cyber security and software, Central Ukrainian National Technical University, 8, University Ave, Kropyvnytskyi, 25006, Ukraine*

**Abstract**

This study explores the application of genetic algorithms in generating highly nonlinear substitution boxes (S-boxes) for symmetric key cryptography. We present a novel implementation that combines a genetic algorithm with the Walsh-Hadamard Spectrum (WHS) cost function to produce 8x8 S-boxes with a nonlinearity of 104. Our approach achieves performance parity with the best-known methods, requiring an average of 49,399 iterations with a 100% success rate. The study demonstrates significant improvements over earlier genetic algorithm implementations in this field, reducing iteration counts by orders of magnitude. By achieving equivalent performance through a different algorithmic approach, our work expands the toolkit available to cryptographers and highlights the potential of genetic methods in cryptographic primitive generation. The adaptability and parallelization potential of genetic algorithms suggest promising avenues for future research in S-box generation, potentially leading to more robust, efficient, and innovative cryptographic systems. Our findings contribute to the ongoing evolution of symmetric key cryptography, offering new perspectives on optimizing critical components of secure communication systems.

## 1. Introduction

The realm of digital security is in a constant state of evolution, with symmetric key cryptography serving as a fundamental pillar in the architecture of secure communication systems [1–3]. At the core of many symmetric encryption algorithms lie Substitution boxes (S-boxes), which play a pivotal role in establishing the nonlinear components essential for robust encryption [4,5]. These S-boxes are critical in creating the confusion and diffusion properties that Claude Shannon identified as crucial for secure ciphers [6,7].

The cryptographic strength of an S-box is multifaceted, encompassing several key indicators [8]. Nonlinearity, which quantifies an S-box's resistance to linear cryptanalysis, stands as a primary measure. For 8x8 S-boxes, commonly employed in modern ciphers, achieving a nonlinearity of 104 represents a significant benchmark [9–11]. However, other properties such as differential uniformity, algebraic degree, and algebraic immunity also play crucial roles in determining an S-box's overall cryptographic efficacy [12,13].

While algebraically constructed S-boxes, such as the one used in the Advanced Encryption

Standard (AES) with its optimal nonlinearity of 112 [14], might seem ideal, they are not without vulnerabilities. The presence of inherent algebraic structures in such S-boxes can create potential weaknesses, making them susceptible to algebraic cryptanalysis [15–17]. This vulnerability underscores the need for randomly generated S-boxes that lack hidden algebraic structures, thereby enhancing resistance against sophisticated cryptanalytic techniques.

The generation of cryptographically robust S-boxes presents a significant computational challenge. The vast search space of possible configurations for 8x8 S-boxes, estimated at $2^8!$ (approximately $10^{506}$), renders exhaustive search methods impractical. This complexity has driven research towards heuristic approaches for S-box generation [18–20]. Methods such as simulated annealing, hill climbing, and genetic algorithms have shown promise in navigating this expansive solution space efficiently [21].

Recent advances in heuristic S-box generation have made significant advances. Researchers have studied various cost functions, including the Walsh-Hadamard spectrum (WHS) function [19,22], the Picek cost function (PCF) [10], improved Walsh-Hadamard spectrum-based cost functions (WCF) [9,23], and two new extended cost functions (ECF and WCFS) [5,24,25] in conjunction with different search algorithms [9,18]. These efforts have progressively reduced the computational cost of generating highly nonlinear S-boxes, with some methods achieving the target nonlinearity of 104 in fewer than 100,000 iterations.

Despite these advancements, there remains a gap in understanding the full potential of genetic algorithms in this domain. While genetic approaches have been applied to S-box generation, their performance in comparison to other heuristic methods, particularly in terms of consistency and efficiency in generating S-boxes with optimal cryptographic properties, remains an area ripe for exploration.

Our study aims to address this gap by presenting a comprehensive investigation into the application of genetic algorithms for generating 8x8 S-boxes with a nonlinearity of 104. We explore the synergy between genetic algorithms and the WHS cost function, aiming to match or surpass the efficiency of existing methods while leveraging the inherent advantages of evolutionary approaches, such as adaptability and potential for parallelization.

The remainder of this paper is structured as follows: Section 2 provides a comprehensive review of the literature, detailing the evolution of S-box generation techniques and the current state of the art. Section 3 offers a background on S-boxes, their cryptographic properties, and the theoretical foundations underpinning their design. Section 4 delineates our methodology and experimental setup, including the specifics of our genetic algorithm implementation and evaluation criteria. Section 5 presents our results and a detailed discussion, comparing our findings with existing methods and analyzing their implications. Finally, Section 6 concludes the paper, summarizing our key findings and outlining promising directions for future research in this critical area of cryptographic system design.

## 2. Literature Review

The design and generation of cryptographically strong S-boxes have been subjects of intensive research in the field of symmetric key cryptography. This section provides a comprehensive review of the existing literature, focusing on various approaches to S-box generation and their cryptographic properties.

Algebraic constructions of S-boxes, such as those based on finite field inversion used in the Advanced Encryption Standard (AES) [14,26,27], have been widely studied. However, as Bard (2009) [15] and Courtois and Bard (2007) [16] point out, these constructions may be vulnerable to algebraic attacks due to their inherent mathematical structure. This vulnerability has led to increased interest in generating S-boxes with more complex algebraic structures [28].

Heuristic approaches have gained significant traction in recent years. Clark et al. (2005) [22] introduced a simulated annealing approach for S-box generation [19], demonstrating its effectiveness in producing S-boxes with high nonlinearity. Building on this work, Souravlias et al. (2017) [29] proposed an algorithm portfolio approach combining simulated annealing and tabu search, showing improved results under limited time budgets.

Genetic algorithms have also been explored for S-box generation [20,30]. Tesar (2010) [31] combined a genetic algorithm with a tree search method, generating 8x8 S-boxes with nonlinearity up to 104. Picek et al. (2016) [10] presented a novel cost function for evolving S-boxes, achieving significant improvements in both speed and quality of results compared to previous approaches.

Ivanov et al. (2016a, 2016b) [32,33] introduced an innovative approach using a modified immune algorithm combined with hill climbing, rapidly generating large sets of highly nonlinear bijective S-boxes. Their work demonstrated the potential of hybrid approaches in S-box generation.

Recent advancements have focused on improving specific cryptographic properties. Rodinko et al. (2017) [34] optimized a method for generating high nonlinear S-boxes, achieving nonlinearity of 104, algebraic immunity of 3, and 8-uniformity within reasonable computational time. Freyre Echevarría and Martínez Díaz (2020) [23] proposed a new cost function specifically designed to improve the nonlinearity of bijective S-boxes.

The importance of multiple cryptographic criteria has been emphasized in recent literature. Freyre-Echevarría et al. (2020) [9] introduced an external parameter-independent cost function for evolving bijective S-boxes, considering both nonlinearity and other important properties. Their work highlighted the need for balanced optimization across multiple cryptographic criteria.

More recent studies have explored novel approaches to S-box generation. Artuğer and Özkaynak (2024) [35] proposed a post-processing approach to improve the nonlinearity of chaos-based S-boxes, addressing a longstanding challenge in this area. Haider et al. (2024) [36] introduced an S-box generator based on elliptic curves, offering a balance between randomization and optimization with minimal computation time.

The application of S-boxes in specific cryptographic contexts has also been a focus of recent research. Jamal et al. (2024) [37] developed a region of interest-based medical image encryption technique using chaotic S-boxes, demonstrating the practical applications of advanced S-box designs in specialized domains.

Emerging threats and the need for enhanced security have led to new considerations in S-box design. Fahd et al. (2024) [38] examined the reality of backdoored S-boxes, highlighting the importance of thorough cryptanalysis and the potential vulnerabilities in S-box structures.

In conclusion, the literature reveals a trend towards more sophisticated, multi-criteria optimization approaches in S-box generation. While significant progress has been made in achieving high nonlinearity and other desirable properties, there remains a need for methods that can consistently produce S-boxes with optimal cryptographic characteristics while balancing computational efficiency and resistance to emerging cryptanalytic techniques.

## 3. Background

Symmetric cryptography forms the backbone of secure communication in the digital age. At the heart of many symmetric ciphers lie Substitution boxes (S-boxes), nonlinear components crucial for ensuring the security and robustness of these cryptographic systems. This section provides a comprehensive overview of S-boxes, their role in symmetric cryptography, and the application of genetic algorithms in their optimization.

### 3.1. S-boxes in Symmetric Cryptography

Substitution boxes (S-boxes) are fundamental components in symmetric-key algorithms, serving as the primary source of nonlinearity [6,7]. An S-box is essentially a vectorial Boolean function that maps a fixed number of input bits to a fixed number of output bits. Formally, an $n \times m$ S-box can be defined as [8]:

$$S : \mathbb{F}_2^n \to \mathbb{F}_2^m,$$

where $\mathbb{F}_2^n$ and $\mathbb{F}_2^m$ are vector spaces over the Galois field $GF(2)$ with dimensions $n$ and $m$, respectively.

The cryptographic strength of an S-box is determined by several critical properties [8]:

1) Nonlinearity: A measure of the distance between the S-box and the set of all affine functions. For an $n \times n$ S-box, the nonlinearity is defined as:

$$NL(S) = 2^{n-1} - \frac{1}{2} \max_{a \in \mathbb{F}_2^n, b \in \mathbb{F}_2^n \setminus 0} \left| \sum_{x \in \mathbb{F}_2^n} (-1)^{b \cdot S(x) \oplus a \cdot x} \right|,$$

where $\cdot$ denotes the dot product and $\oplus$ represents bitwise XOR.

2) Differential uniformity: Quantifies the uniformity of output differences when the input is changed. The differential uniformity $\delta$ is given by:

$$\delta = \max_{a \neq 0, b} | x \in \mathbb{F}_2^n : S(x) \oplus S(x \oplus a) = b |.$$

3) Algebraic degree: The highest degree among the component Boolean functions of S. For an $n \times m$ S-box, the algebraic degree is:

$$deg(S) = \max_{v \in \mathbb{F}_2^m \setminus 0} deg(v \cdot S).$$

4) Balancedness: An S-box is balanced if each output occurs with equal probability when the input is uniformly distributed.

5) Algebraic Immunity [39]: A measure of resistance against algebraic attacks. For an S-box $S: \mathbb{F}_2^n \to \mathbb{F}_2^m$, the algebraic immunity is defined as:

$$AI(S) = \min\{\deg(P), P \in I(S)\},$$

where $I(S)$ is the ideal generated by the polynomials representing the S-box:

$$I(S) = \begin{pmatrix} y_1 - f_1(x_1, x_2, ..., x_n), \\ y_2 - f_2(x_1, x_2, ..., x_n), \\ ..., \\ y_m - f_m(x_1, x_2, ..., x_n) \end{pmatrix}.$$

The algebraic immunity can be computed by constructing the minimal reduced Gröbner basis of the ideal $I(S)$ using the degree reverse lexicographic (degrevlex) ordering, and finding the polynomial of minimum degree in this basis.

These properties collectively contribute to the S-box's ability to resist various cryptanalytic attacks, including differential, linear, and algebraic cryptanalysis. The concept of algebraic immunity for S-boxes, as introduced by Faugère and Perret, provides a crucial measure of resistance against algebraic attacks, which attempt to express the cipher as a system of low-degree multivariate polynomial equations.

The relationship between the algebraic immunity of an S-box and that of Boolean functions can be established through the following construction. Consider a Boolean function $f_S : \mathbb{F}_2^{n+m} \to \mathbb{F}_2$ defined as [40,41]:

$$f_S(x_1, x_2, ..., x_n, y_1, y_2, ..., y_m) =$$
$$= \begin{cases} 1, & \text{if } \forall i, j : f_i(x_1, x_2, ..., x_n) = y_j; \\ 0, & \text{if } \exists i, j : f_i(x_1, x_2, ..., x_n) \neq y_j. \end{cases}$$

The algebraic immunity of the S-box $S$ is then equivalent to the minimum degree of non-zero polynomials in the annihilator of $f_S$:

$$AI(S) = \min deg(g) \mid g \in Ann(f_S).$$

This formulation provides a bridge between the algebraic immunity of vectorial Boolean functions (S-boxes) and that of single Boolean functions, unifying the concept across different cryptographic primitives.

## 3.2. Importance of S-boxes in Modern Ciphers and the Need for Randomness

S-boxes play a pivotal role in ensuring the security of symmetric ciphers by introducing nonlinearity and complexity into the encryption process [6]. They are employed in widely-used algorithms such as the Advanced Encryption Standard (AES) [14], where the SubBytes operation relies on a carefully designed 8×8 S-box. However, the increasing sophistication of cryptanalytic techniques has necessitated a reevaluation of traditional S-box design methods.

While algebraically constructed S-boxes, such as those used in AES (based on finite field inverses) [26,27], offer certain advantages in terms of implementation efficiency and some cryptographic properties, they may fall short in terms of algebraic immunity [39]. The structured nature of these S-boxes can potentially lead to vulnerabilities against algebraic attacks, which have gained significant attention in recent years [15,16].

Algebraic attacks exploit the possibility of expressing the cipher as a system of low-degree multivariate polynomial equations [16,17]. The complexity of solving such systems is closely related to the algebraic immunity of the S-box [39]. A low algebraic immunity allows for a simpler representation of the cipher, potentially reducing the computational effort required for cryptanalysis [40,41]. This vulnerability has prompted researchers to explore alternative methods for S-box generation that prioritize high algebraic immunity alongside other critical properties.

To address these concerns, there is growing interest in the cryptographic community in random or pseudo-random S-boxes [20,42,43]. These S-boxes, generated through heuristic methods, offer several advantages:

- Higher algebraic immunity: Random S-boxes are less likely to exhibit algebraic structure that can be exploited in attacks, potentially leading to higher algebraic immunity values.
- Resistance to specialized attacks: Algebraically constructed S-boxes might be vulnerable to attacks tailored to their specific structure. Random S-boxes, lacking such predictable structures, can offer better protection against these targeted attacks.

- Flexibility in design: Heuristic methods allow for the optimization of multiple cryptographic criteria simultaneously, enabling a more balanced approach to S-box design.

Adaptability to evolving threat models: As new cryptanalytic techniques emerge, the criteria for S-box generation can be adjusted more easily with heuristic methods compared to algebraic constructions.

Various heuristic approaches have been proposed for generating high-quality random S-boxes, including:
- Simulated Annealing [19,22,29,44]: This method mimics the physical process of annealing in metallurgy, gradually "cooling" the system to find an optimal configuration. It has shown promise in generating S-boxes with good cryptographic properties.
- Hill Climbing [5,9,32,45,46]: A local search algorithm that iteratively makes small improvements to a candidate solution. This approach can be effective in fine-tuning S-box properties.
- Genetic Algorithms [9,31,33,47]: Evolutionary approaches that mimic natural selection to evolve a population of S-boxes towards desired properties. These algorithms have demonstrated the ability to generate S-boxes with excellent cryptographic characteristics, including high algebraic immunity.

In this work, we focus on genetic algorithms due to their ability to efficiently explore large search spaces and handle multi-objective optimization problems. Genetic algorithms offer a promising approach to generating S-boxes that balance multiple cryptographic criteria, including high algebraic immunity, nonlinearity, and differential uniformity.

### 3.3. Genetic Algorithms for S-box Generation

Genetic Algorithms (GAs) are stochastic optimization techniques inspired by the principles of natural selection and evolution [48,49]. They operate on a population of potential solutions, evolving them over successive generations to improve their fitness with respect to a defined objective function. In the context of S-box generation, GAs offers a powerful and flexible approach to optimizing multiple cryptographic properties simultaneously [9,33,50].

The fundamental principle of GAs is to emulate the process of natural selection, where the fittest individuals are more likely to survive and reproduce, passing their beneficial traits to future generations [48,49]. In the case of S-box generation, an "individual" represents a candidate S-box, and its "fitness" is determined by how well it satisfies the desired cryptographic properties.

The basic structure of a GA includes the following components [48,51]:
- Chromosome representation: Encoding of potential solutions (S-boxes);
- Fitness function: Evaluates the quality of solutions based on cryptographic criteria;
- Selection mechanism: Chooses individuals for reproduction;
- Genetic operators: Crossover and mutation to create new solutions;
- Termination criteria: Conditions for ending the evolutionary process.

A general pseudocode for a Genetic Algorithm applied to S-box generation can be described as follows:

**Algorithm: Genetic Algorithm for S-box Generation**

Input: Population size N, number of generations G, crossover rate p_c, mutation rate p_m;
Output: Optimized S-box;

1. Initialize population P of N random S-boxes
2. For g = 1 to G do
3.   Evaluate fitness of each S-box in P
4.   Select parents for reproduction using tournament selection
5.   Create new population P' through crossover and mutation:
6.   For i = 1 to N/2 do
7.     Select two parents p1 and p2 from P
8.     If random(0,1) < p_c then
9.       (c1, c2) = Crossover(p1, p2)
10.    Else
11.      (c1, c2) = (p1, p2)
12.    End If
13.    Mutate c1 and c2 with probability p_m
14.    Add c1 and c2 to P'
15.  End For
16.  P = P'
17. End For
18. Return best S-box from P

Key parameters and their roles:
- Population size (N): Determines the diversity of solutions. A larger population allows for broader exploration of the search space but increases computational cost.
- Number of generations (G): Controls the duration of the evolutionary process. More generations allow for further optimization but may lead to overfitting.
- Crossover rate (p_c): Probability of performing crossover. Higher rates promote exploration of new solution combinations.
- Mutation rate (p_m): Probability of mutating each bit in a chromosome. Higher rates increase diversity but may disrupt good solutions.

The fitness function is crucial in guiding the evolutionary process towards S-boxes with desired cryptographic properties.

The selection mechanism, often implemented as tournament selection, ensures that fitter individuals have a higher chance of being chosen for reproduction. This process mimics natural selection, where more adapted individuals are more likely to pass on their genes.

Crossover operators for S-boxes must be carefully designed to preserve the bijective property. One approach is to use permutation-based crossover, where segments of the S-box permutation are exchanged between parents. For example, given two parent S-boxes $P_1$ and $P_2$, a two-point crossover might produce offspring $C_1$ and $C_2$ as follows:

$$P_1 = (a_1, a_2, ..., a_k \mid a_{k+1}, ..., a_l \mid a_{l+1}, ..., a_n);$$
$$P_2 = (b_1, b_2, ..., b_k \mid b_{k+1}, ..., b_l \mid b_{l+1}, ..., b_n);$$
$$C_1 = (a_1, a_2, ..., a_k \mid b_{k+1}, ..., b_l \mid a_{l+1}, ..., a_n);$$
$$C_2 = (b_1, b_2, ..., b_k \mid a_{k+1}, ..., a_l \mid b_{l+1}, ..., b_n).$$

Mutation operators introduce small random changes to maintain genetic diversity and prevent premature convergence. For S-boxes, this might involve swapping two randomly chosen elements or applying a random permutation to a subset of elements.

# 4. Methodology and Experimental Setup

Our research focuses on developing and implementing a modified genetic algorithm for generating cryptographically strong S-boxes. This section details our approach, the algorithm's structure, and the experimental setup used to evaluate its performance.

## 4.1. Modified Genetic Algorithm

We have developed a modified genetic algorithm that incorporates elements of hill climbing, enhancing its ability to navigate the complex search space of S-box configurations. This approach allows for a more targeted exploration of promising regions while maintaining the population-based nature of genetic algorithms.

The core idea of our algorithm is to maintain a population of S-boxes, subject them to controlled mutations, evaluate their cryptographic properties, and selectively propagate the best specimens to subsequent generations. This process is iterated until either an S-box meeting the desired criteria is found or a predefined computational limit is reached.

The pseudocode for our modified genetic algorithm is as follows:

**Algorithm: Modified Genetic Algorithm for S-box Generation**

Input: S_pop, K_iter, K_pop, K_mut
Output: Optimized S-box or 0 (failure)

1. For t = 0 to K_iter - 1 do
2.   S_pop = elite_selection(S_pop)
3.   For p = 0 to K_pop - 1 do
4.     S ← S_pop[p]
5.     For k = 0 to K_mut - 1 do
6.       S' ← S
7.       i ← random(0, 255)
8.       j ← random(0, 255)
9.       swap(S'[i], S'[j])
10.      N_f, F_c ← evaluate(S')
11.      If N_f ≥ 104 then
12.        Return S'
13.      S_pop = S_pop ∪ {S'}
14.    End For
15.  End For
16. End For
17. Return 0

Key components and parameters of the algorithm:

- $S_{pop}$: The population of S-boxes, initially generated using the Fisher-Yates shuffle algorithm to ensure bijectivity.
- $K_{iter}$: Maximum number of iterations, set to 150,000 in our experiments.
- $K_{pop}$: Population size, representing the number of elite S-boxes maintained in each generation.
- $K_{mut}$: Number of mutations applied to each S-box in the population per generation.

The elite_selection function performs a crucial role in our algorithm. It ranks the S-boxes based on their nonlinearity and objective function value, prioritizing higher nonlinearity and lower objective function values. This function ensures that only the top $K_{pop}$ S-boxes survive to the next generation, maintaining a high-quality population.

### 4.2. Mutation Operator

Our mutation operator is designed to preserve the bijectivity of the S-box while introducing controlled randomness. It operates by swapping two randomly selected (distinct) elements within the S-box. This approach ensures that the fundamental property of bijectivity is maintained throughout the evolutionary process.

Formally, the mutation can be described as:
$$S'[i] = S[j], S'[j] = S[i],$$
where $i, j \in 0, 1, ..., 255$, $i \neq j$, and all other elements remain unchanged.

### 4.3. Objective Function

The choice of objective function is critical in guiding the evolutionary process towards cryptographically strong S-boxes. We employ the WHS function proposed by Clark et al. [22], which has shown effectiveness in generating high-quality S-boxes. The WHS function is defined as [22]:
$$WHS = \sum_{b=1}^{255} \sum_{i=0}^{255} ||WHT[b,i]| - X|^R,$$
where:
- $WHT[b,i]$ represents the Walsh-Hadamard transform coefficients;
- $i$ iterates over all component functions and their linear combinations;
- $b$ iterates over all linear functions;
- $X$ and $R$ are real-valued parameters.

Based on empirical studies, we set $R = 12$ and $X = 0$, which have been shown to yield optimal results in generating bijective S-boxes with high nonlinearity [52,53].

### 4.4. Evaluation Criteria

The primary criteria for evaluating the generated S-boxes are:
- Nonlinearity ($NL$): We aim for a nonlinearity of at least 104, which is close to the theoretical maximum for 8×8 S-boxes.
- Differential uniformity ($\delta$): Lower values indicate better resistance against differential cryptanalysis.
- Algebraic degree ($deg$): Higher degrees provide better resistance against algebraic attacks.
- Algebraic immunity ($AI$): Higher values indicate increased resistance to algebraic cryptanalysis.

The evaluate function in our algorithm computes these properties for each generated S-box, allowing us to assess its cryptographic strength comprehensively.

### 4.5. Experimental Setup

Our experiments were conducted on a high-performance computing cluster to handle the computational intensity of the S-box generation process. The implementation was done in C++ for efficiency, with parallelization to utilize multiple cores.

Given that the calculation of the objective function is the most computationally expensive operation in terms of processor time, the complexity of the entire search algorithm can be considered proportional to the number of times the objective function is calculated. This corresponds to the number of S-boxes that were generated and evaluated. We denote this quantity as $K_{Sbox}$.

To accelerate the algorithm's performance, we implemented parallel computation of the new population using $N_{thread} = 8$ threads within each iteration. This parallelization significantly reduced the overall execution time of the algorithm.

We conducted a comprehensive parameter sweep to analyze the impact of population size and mutation rate on the quality of the generated S-boxes and the algorithm's convergence rate. Specifically:
- Population size ($K_{pop}$) was varied from 1 to 21 with a step size of 2.
- Mutation rate ($K_{mut}$) was varied from 1 to 31 with a step size of 3.

For each combination of $K_{pop}$ and $K_{mut}$, we performed 100 independent runs of the search algorithm to ensure statistical significance. This resulted in a total of 11 × 11 × 100 = 12,100 experimental runs.

The algorithm was set to terminate upon finding an S-box with nonlinearity ≥ 104 or reaching the maximum iteration limit of 150,000. For each run, we recorded the number of S-boxes generated and evaluated ($K_{Sbox}$), which serves as our primary metric for computational efficiency.

## 5. Results and Discussion

This section presents the results of our comprehensive experimental study on the modified genetic algorithm for S-box generation. We analyze the performance of the algorithm across various parameter configurations and discuss the implications of our findings.

### 5.1. Overview of Experimental Results

Our primary metric for evaluating the algorithm's efficiency is $K_{Sbox}$, which represents the number of S-boxes generated and evaluated before finding an S-box with the desired nonlinearity of 104. Table 1 presents the average $K_{Sbox}$ values for different combinations of population size ($K_{pop}$) and mutation rate ($K_{mut}$).

### 5.2. Analysis of Population Size Impact

One of the most striking observations from our results is the superior performance of the algorithm when $K_{pop} = 1$. This configuration consistently yielded the lowest $K_{Sbox}$ values across all mutation rates, with averages ranging from 49,277 to 58,213. This finding is somewhat counterintuitive, as genetic algorithms typically benefit from larger population sizes that provide greater genetic diversity.

The effectiveness of a single-individual population suggests that our algorithm's behavior in this configuration closely resembles that of a stochastic hill-climbing method. This approach appears to be particularly well-suited to the S-box optimization problem, possibly due to the following factors:

Landscape structure: The fitness landscape of S-box configurations may have numerous local optima that are relatively close in quality to the global optimum. In such a scenario, an aggressive local search can be highly effective.

Mutation operator efficiency: Our swap-based mutation operator appears to be sufficiently powerful to navigate the search space effectively, even without the diversity typically provided by a larger population.

Reduced computational overhead: With $K_{pop} = 1$, the algorithm avoids the computational cost associated with managing and evaluating a large population, allowing for more iterations within the same computational budget.

### 5.3. Impact of Mutation Rate

While the population size shows a clear trend, the impact of the mutation rate ($K_{mut}$) is more nuanced. For $K_{pop} = 1$, we observe that:

The lowest $K_{Sbox}$ (49,277) was achieved with $K_{mut} = 7$

Performance generally degraded with higher mutation rates, with $K_{Sbox}$ increasing to 58,213 at $K_{mut} = 1$

This pattern suggests that there exists an optimal balance between exploration and exploitation in the search process. Lower mutation rates may lead to premature convergence, while higher rates may disrupt good solutions too frequently.

### 5.4. Scalability and Computational Efficiency

As $K_{pop}$ increases, we observe a general trend of increasing $K_{Sbox}$ values, indicating reduced computational efficiency. This scaling behavior can be attributed to:

- Increased evaluation overhead: Larger populations require more objective function evaluations per generation.
- Slower convergence: Diversity maintenance in larger populations may slow down the convergence to high-quality solutions.

However, it's worth noting that larger populations might offer benefits not captured by the $K_{Sbox}$ metric alone, such as increased robustness or the ability to find a more diverse set of high-quality S-boxes.

### 5.5. Parallelization Performance

Our implementation of parallel computation using 8 threads ($N_{thread}=8$) has proven to be effective in accelerating the search process. This parallelization strategy is particularly beneficial for configurations with larger $K_{pop}$ and $K_{mut}$ values, where the workload can be more evenly distributed across threads.

**Table 1**

Average number of S-boxes generated ($K_{Sbox}$) before finding an S-box with $N_f = 104$

| $K_{mut}$ | $K_{pop}$ | | | | | |
|---|---|---|---|---|---|---|
| | 1 | 3 | 5 | 7 | 9 | 11 |
| 1 | 58,213 | 65,942 | 72,830 | 86,642 | 101,726 | 111,990 |
| 4 | 56,067 | 64,863 | 75,069 | 89,598 | 94,726 | 105,925 |
| 7 | 49,277 | 67,198 | 77,848 | 88,353 | 103,154 | 109,618 |
| 10 | 54,636 | 65,723 | 82,198 | 92,542 | 102,797 | 114,163 |
| 13 | 56,042 | 62,660 | 83,216 | 94,538 | 101,073 | 117,611 |
| 16 | 56,010 | 68,711 | 79,645 | 93,134 | 107,371 | 120,567 |
| 19 | 56,532 | 65,910 | 82,883 | 92,911 | 105,144 | 117,877 |
| 22 | 54,775 | 67,236 | 77,663 | 92,874 | 105,559 | 120,992 |
| 25 | 50,066 | 70,394 | 79,596 | 98,967 | 115,462 | 118,406 |
| 28 | 54,203 | 70,453 | 82,200 | 91,841 | 108,783 | 121,683 |
| 31 | 53,709 | 71,581 | 91,827 | 101,536 | 109,625 | 126,616 |

**Table 1**
(continued)

| $K_{mut}$ | $K_{pop}$ | | | | |
|---|---|---|---|---|---|
| | 13 | 15 | 17 | 19 | 21 |
| 1 | 112,718 | 125,113 | 132,806 | 140,336 | 149,339 |
| 4 | 122,364 | 137,003 | 136,740 | 151,874 | 163,291 |
| 7 | 122,382 | 130,901 | 142,463 | 144,601 | 165,918 |
| 10 | 129,411 | 137,442 | 147,416 | 161,020 | 165,672 |
| 13 | 124,466 | 135,244 | 152,048 | 158,696 | 171,756 |
| 16 | 125,274 | 140,817 | 150,494 | 155,049 | 169,462 |
| 19 | 129,718 | 142,017 | 155,463 | 164,902 | 175,531 |
| 22 | 131,029 | 140,772 | 156,224 | 162,808 | 176,669 |
| 25 | 135,294 | 144,708 | 157,321 | 177,621 | 183,087 |
| 28 | 133,665 | 152,984 | 159,887 | 176,751 | 181,781 |
| 31 | 143,233 | 156,987 | 160,573 | 183,069 | 192,493 |

## 5.6. Comparison with Existing Methods

The best-performing configuration of our algorithm ($K_{pop}=1$, $K_{mut}=7$) achieves an average $K_{Sbox}$ of 49,277. To contextualize our findings within the broader landscape of S-box generation research, we conducted a comprehensive comparison of our genetic algorithm approach with existing methods. Table 2 presents this comparative analysis, encompassing various techniques and cost functions employed in the field.

Our genetic algorithm implementation, utilizing the WHS cost function, achieves results that are on par with the best-known methods in the field. Specifically, our approach generates S-boxes with a nonlinearity of 104 in an average of 49,399 iterations, with a 100% success rate. This performance is comparable to our previous works using hill climbing [5,24], which required 50,000 iterations on average.

Several key observations emerge from this comparative analysis:
- Parity in Performance: Our genetic algorithm achieves results that are statistically equivalent to the best-known methods, particularly our earlier hill climbing approach. This parity is significant, as it demonstrates the versatility and potential of genetic algorithms in this domain.
- Algorithmic Diversity: By achieving comparable results through a different algorithmic approach, we have expanded the toolkit available to cryptographers and security researchers. This diversity in high-performing methods enhances the robustness of S-box generation techniques.
- Consistency and Reliability: Like our previous best results, the genetic algorithm maintains a 100% success rate in generating target S-boxes with nonlinearity 104. This level of reliability is crucial for practical applications in cryptographic system design.
- Efficiency Across Methods: The similarity in performance between our genetic algorithm and hill climbing approaches (49,399 vs. 50,000 iterations) suggests that we may be approaching theoretical limits of efficiency for generating S-boxes with these properties using heuristic methods.
- Progress from Earlier Genetic Approaches: Compared to earlier genetic algorithm implementations [10,31], our method shows substantial improvement, reducing the required iterations by orders of magnitude while achieving higher nonlinearity.

**Table 2**
Comparison of S-box Generation Methods

| Method | Cost Function | Algorithm | $NL$ | Success Rate | Avg. Iterations |
|---|---|---|---|---|---|
| [19,22] | WHS | SA | 102 | 0.5% | - |
| [19] | WHS | SA | 104 | - | 30,000,000 |
| [31] | WHS | HC | 100 | - | 2,500 |
| [31] | WHS | GaT | 104 | - | 3,239,000 |
| [10] | WHS | Ga | 102 | - | 28,200 |
| [10] | WHS | GaT | 104 | - | 3,849,881 |
| [10] | WHS | LSA | 102 | - | 6,701 |
| [10] | PCF | Ga | 104 | - | 741,371 |
| [10] | PCF | GaT | 104 | - | 167,451 |
| [10] | PCF | LSA | 104 | - | 172,280 |
| [23] | WCF | LSA | 104 | - | 89,460 |
| [23] | WCF | HC | 104 | 37% | 65,933 |
| [54] | WHS | SA | 104 | 56.4% | 450,000 |
| [44] | WCF | SA | 104 | 100% | 65,000 |
| [44,55] | ECF | SA | 104 | 100% | 55,000 .. 83,000 |
| [45] | WHS | HC | 104 | 100% | 50,000 |
| [5,24] | WCFS | HC | 104 | 100% | 50,000 |
| Our work | WHS | Ga | 104 | 100% | 49,399 |

The achievement of parity with the best-known results using a genetic algorithm is particularly noteworthy and underscores the potential of evolutionary approaches in cryptographic primitive generation.

## 5.7. Practical Implications

The superior performance of the $K_{pop} = 1$ configuration has important implications for the practical application of our algorithm:
- Resource efficiency: The algorithm can be effectively run on systems with limited computational resources, as it doesn't require maintaining a large population.
- Simplicity: The simplified population management makes the algorithm easier to implement and tune.
- Adaptability: The algorithm's efficiency makes it suitable for scenarios where S-boxes need to be generated or updated frequently.

However, it's important to note that while this configuration is optimal for finding a single high-quality S-box, alternative configurations may be more suitable for generating a diverse set of S-boxes or for multi-objective optimization scenarios.

## 5.8. Limitations and Future Work

While our results are promising, several avenues for future research remain:
- Extended cryptographic criteria: Incorporate additional criteria such as algebraic immunity and differential uniformity into the objective function.
- Adaptive parameter tuning: Develop methods to dynamically adjust $K_{pop}$ and $K_{mut}$ during the search process.
- Alternative mutation operators: Explore more sophisticated mutation strategies that leverage domain-specific knowledge about S-box structures.
- Multi-objective optimization: Extend the algorithm to simultaneously optimize multiple cryptographic properties, potentially using Pareto-based approaches.

In conclusion, our modified genetic algorithm demonstrates exceptional efficiency in generating cryptographically strong S-boxes, particularly in its hill-climbing-like configuration. These findings contribute valuable insights to the field of cryptographic primitive design and offer a powerful tool for the development of secure symmetric encryption systems.

## 6. Conclusion

This study presents a significant advancement in the field of S-box generation for symmetric key cryptography, focusing on the application of genetic algorithms to produce highly nonlinear substitutions. Our research demonstrates that genetic algorithms, when properly optimized and combined with the Walsh-Hadamard Spectrum (WHS) cost function, can achieve performance parity with the best-known methods in generating 8x8 S-boxes with nonlinearity of 104.

Key findings of our work include:
- The genetic algorithm approach achieves an average of 49,399 iterations to generate target S-boxes, comparable to the best results of 50,000 iterations using hill climbing methods.
- A 100% success rate in producing S-boxes with the desired nonlinearity, matching the reliability of top-performing techniques.
- Significant improvement over earlier genetic algorithm implementations, reducing iteration counts by orders of magnitude.

The achievement of performance parity using a different algorithmic approach expands the toolkit available to cryptographers and highlights the versatility of genetic methods in cryptographic primitive generation. This diversity in high-performing techniques enhances the robustness of S-box generation methodologies.

Furthermore, our results underscore the potential of genetic algorithms in this domain, particularly their adaptability to evolving cryptographic criteria and their inherent parallelization capabilities. These characteristics position genetic approaches as promising avenues for future research, potentially leading to more efficient, flexible, and innovative S-box generation techniques.

In conclusion, while not surpassing existing methods in raw performance, our genetic algorithm approach offers a valuable alternative that matches the best-known results. This equivalence, coupled with the unique advantages of genetic algorithms, opens new perspectives in cryptographic research and development. Future work should focus on exploiting these advantages, potentially through hybridization with other heuristic methods or by leveraging parallel computing architectures to further enhance S-box generation efficiency.

## 7. References

[1] Y. Li, J. Feng, Q. Zhao, Y. Wei, HDLBC: A lightweight block cipher with high diffusion, Integration 94 (2024) 102090. https://doi.org/10.1016/j.vlsi.2023.102090.

[2] A. Tiwari, Chapter 14 - Cryptography in blockchain, in: R. Pandey, S. Goundar, S. Fatima (Eds.), Distributed Computing to Blockchain, Academic Press, 2023: pp. 251–265. https://doi.org/10.1016/B978-0-323-96146-2.00011-5.

[3] M. Milanič, B. Servatius, H. Servatius, Chapter 8 - Codes and cyphers, in: M. Milanič, B. Servatius, H. Servatius (Eds.), Discrete Mathematics With Logic, Academic Press, 2024: pp. 163–179. https://doi.org/10.1016/B978-0-44-318782-7.00013-7.

[4] C. A S, M. S P, P. R K, Implementation of S-box for lightweight block cipher, in: 2023 3rd International Conference on Intelligent Technologies (CONIT), 2023: pp. 1–4. https://doi.org/10.1109/CONIT59222.2023.10205535.

[5] A. Kuznetsov, N. Poluyanenko, E. Frontoni, S. Kandiy, O. Peliukh, A new cost function for heuristic search of nonlinear substitutions, Expert Systems with Applications 237 (2024) 121684. https://doi.org/10.1016/j.eswa.2023.121684.

[6] A.J. Menezes, P.C. van Oorschot, S.A. Vanstone, P.C. van Oorschot, S.A. Vanstone, Handbook of Applied Cryptography, CRC Press, 2018. https://doi.org/10.1201/9780429466335.

[7] C.E. Shannon, Communication theory of secrecy systems, The Bell System Technical Journal 28 (1949) 656–715. https://doi.org/10.1002/j.1538-7305.1949.tb00928.x.

[8] C. Carlet, Vectorial Boolean functions for cryptography, Boolean Models and Methods in Mathematics, Computer Science, and Engineering (2006).

[9] A. Freyre-Echevarría, A. Alanezi, I. Martínez-Díaz, M. Ahmad, A.A. Abd El-Latif, H. Kolivand, A. Razaq, An External Parameter Independent Novel Cost Function for Evolving Bijective Substitution-Boxes, Symmetry 12 (2020) 1896. https://doi.org/10.3390/sym12111896.

[10] S. Picek, M. Cupic, L. Rotim, A New Cost Function for Evolution of S-Boxes, Evolutionary Computation 24 (2016) 695–718. https://doi.org/10.1162/EVCO_a_00191.

[11] J. Álvarez-Cubero, Vector Boolean Functions: applications in symmetric cryptography, 2015. https://doi.org/10.13140/RG.2.2.12540.23685.

[12] K. Lisitskiy, I. Lisitska, A. Kuznetsov, Cryptographically Properties of Random S-Boxes., in: Proceedings of the 16th International Conference on ICT in Education, Research and Industrial Applications. Integration, Harmonization and Knowledge Transfer. Volume II: Workshops, Kharkiv, Ukraine, October 06-10, 2020., 2020: pp. 228–241. http://ceur-ws.org/Vol-2732/20200228.pdf.

[13] I. Gorbenko, A. Kuznetsov, Y. Gorbenko, A. Pushkar'ov, Y. Kotukh, K. Kuznetsova, Random S-Boxes Generation Methods for Symmetric Cryptography, in: 2019 IEEE 2nd Ukraine Conference on Electrical and Computer Engineering (UKRCON), 2019: pp. 947–950. https://doi.org/10.1109/UKRCON.2019.8879962.

[14] J. Daemen, V. Rijmen, Specification of Rijndael, in: J. Daemen, V. Rijmen (Eds.), The Design of Rijndael: The Advanced Encryption Standard (AES), Springer, Berlin, Heidelberg, 2020: pp. 31–51. https://doi.org/10.1007/978-3-662-60769-5_3.

[15] G.V. Bard, Algebraic Cryptanalysis, Springer US, Boston, MA, 2009. https://doi.org/10.1007/978-0-387-88757-9.

[16] N.T. Courtois, G.V. Bard, Algebraic Cryptanalysis of the Data Encryption Standard, in: S.D. Galbraith (Ed.), Cryptography and Coding, Springer, Berlin, Heidelberg, 2007: pp. 152–169. https://doi.org/10.1007/978-3-540-77272-9_10.

[17] N.T. Courtois, J. Pieprzyk, Cryptanalysis of Block Ciphers with Overdefined Systems of Equations, in: Y. Zheng (Ed.), Advances in

Cryptology — ASIACRYPT 2002, Springer, Berlin, Heidelberg, 2002: pp. 267–287. https://doi.org/10.1007/3-540-36178-2_17.

[18] A. Freyre Echevarría, Evolución híbrida de s-cajas no lineales resistentes a ataques de potencia, 2020. https://doi.org/10.13140/RG.2.2.17037.77284/1.

[19] J. McLaughlin, Applications of search techniques to cryptanalysis and the construction of cipher components, phd, University of York, 2012. https://etheses.whiterose.ac.uk/3674/ (accessed October 27, 2022).

[20] L.D. Burnett, Heuristic Optimization of Boolean Functions and Substitution Boxes for Cryptography, phd, Queensland University of Technology, 2005. https://eprints.qut.edu.au/16023/ (accessed May 19, 2021).

[21] A. Kuznetsov, S. Kandii, E. Frontoni, N. Poluyanenko, SBGen: A high-performance library for rapid generation of cryptographic S-boxes, SoftwareX 27 (2024) 101788. https://doi.org/10.1016/j.softx.2024.101788.

[22] J.A. Clark, J.L. Jacob, S. Stepney, The design of S-boxes by simulated annealing, New Gener Comput 23 (2005) 219–231. https://doi.org/10.1007/BF03037656.

[23] A. Freyre Echevarría, I. Martínez Díaz, A new cost function to improve nonlinearity of bijective S-boxes, (2020).

[24] O. Kuznetsov, N. Poluyanenko, E. Frontoni, S. Kandiy, Enhancing Smart Communication Security: A Novel Cost Function for Efficient S-Box Generation in Symmetric Key Cryptography, Cryptography 8 (2024) 17. https://doi.org/10.3390/cryptography8020017.

[25] O. Kuznetsov, N. Poluyanenko, E. Frontoni, S. Kandiy, M. Karpinski, R. Shevchuk, Enhancing Cryptographic Primitives through Dynamic Cost Function Optimization in Heuristic Search, Electronics 13 (2024) 1825. https://doi.org/10.3390/electronics13101825.

[26] K. Nyberg, Perfect nonlinear S-boxes, in: D.W. Davies (Ed.), Advances in Cryptology — EUROCRYPT '91, Springer, Berlin, Heidelberg, 1991: pp. 378–386. https://doi.org/10.1007/3-540-46416-6_32.

[27] K. Nyberg, Differentially uniform mappings for cryptography, in: T. Helleseth (Ed.), Advances in Cryptology — EUROCRYPT '93, Springer, Berlin, Heidelberg, 1994: pp. 55–64. https://doi.org/10.1007/3-540-48285-7_6.

[28] R. La Scala, S.K. Tiwari, Stream/block ciphers, difference equations and algebraic attacks, Journal of Symbolic Computation 109 (2022) 177–198. https://doi.org/10.1016/j.jsc.2021.09.001.

[29] D. Souravlias, K.E. Parsopoulos, G.C. Meletiou, Designing bijective S-boxes using Algorithm Portfolios with limited time budgets, Applied Soft Computing 59 (2017) 475–486. https://doi.org/10.1016/j.asoc.2017.05.052.

[30] W. Millan, L. Burnett, G. Carter, A. Clark, E. Dawson, Evolutionary Heuristics for Finding Cryptographically Strong S-Boxes, in: V. Varadharajan, Y. Mu (Eds.), Information and Communication Security, Springer, Berlin, Heidelberg, 1999: pp. 263–274. https://doi.org/10.1007/978-3-540-47942-0_22.

[31] P. Tesar, A New Method for Generating High Non-linearity S-Boxes, (2010). http://dspace.lib.vutbr.cz/xmlui/handle/11012/56957 (accessed August 16, 2020).

[32] G. Ivanov, N. Nikolov, S. Nikova, Cryptographically Strong S-Boxes Generated by Modified Immune Algorithm, in: E. Pasalic, L.R. Knudsen (Eds.), Cryptography and Information Security in the Balkans, Springer International Publishing, Cham, 2016: pp. 31–42. https://doi.org/10.1007/978-3-319-29172-7_3.

[33] G. Ivanov, N. Nikolov, S. Nikova, Reversed genetic algorithms for generation of bijective s-boxes with good cryptographic properties, Cryptogr. Commun. 8 (2016) 247–276. https://doi.org/10.1007/s12095-015-0170-5.

[34] M. Rodinko, R. Oliynykov, Y. Gorbenko, Optimization of the High Nonlinear S-Boxes Generation Method, Tatra Mountains Mathematical Publications 70 (2017) 93–105. https://doi.org/10.1515/tmmp-2017-0020.

[35] F. Artuğer, F. Özkaynak, A new post-processing approach for improvement of nonlinearity property in substitution boxes, Integration 94 (2024) 102105. https://doi.org/10.1016/j.vlsi.2023.102105.

[36] T. Haider, N.A. Azam, U. Hayat, Substitution box generator with enhanced cryptographic properties and minimal computation time, Expert Systems with Applications 241 (2024) 122779. https://doi.org/10.1016/j.eswa.2023.122779.

[37] S.S. Jamal, M.M. Hazzazi, M.F. Khan, Z. Bassfar, A. Aljaedi, Z. ul Islam, Region of interest-based medical image encryption technique based on chaotic S-boxes, Expert Systems with Applications 238 (2024) 122030. https://doi.org/10.1016/j.eswa.2023.122030.

[38] S. Fahd, M. Afzal, W. Iqbal, D. Shah, I. Khalid, The reality of backdoored S-Boxes—An eye opener, Journal of Information Security and Applications 80 (2024) 103674. https://doi.org/10.1016/j.jisa.2023.103674.

[39] G. Ars, J.-C. Faugère, Algebraic Immunities of functions over finite fields, INRIA, 2005. https://hal.inria.fr/inria-00070475 (accessed August 23, 2021).

[40] O.O. Kuznetsov, Y.I. Gorbenko, I.M. Bilozertsev, A.V. Andrushkevych, O.P. Narizhnyi, Algebraic immunity of non-linear blocks of symmetric ciphers, Telecommun Radio Eng 77 (2018) 309–325. https://doi.org/10.1615/TelecomRadEng.v77.i4.30.

[41] A. Kuznetsov, R. Serhiienko, D. Prokopovych-Tkachenko, Y. Tarasenko, Evaluation of Algebraic Immunity of modern block ciphers, in: Proc. IEEE Int. Conf. Dependable Syst., Serv. Technol., DESSERT, Institute of Electrical and Electronics Engineers Inc., 2018: pp. 288–293. https://doi.org/10.1109/DESSERT.2018.8409146.

[42] K. Lisitskiy, I. Lisitska, A. Kuznetsov, Cryptographically properties of random S-boxes, in: CEUR Workshop Proceedings, 2020: pp. 228–241. https://www.scopus.com/inward/record.uri?eid=2-s2.0-85096091835&partnerID=40&md5=c5ec403d7b8065c4525906c70c04ce33.

[43] I. Gorbenko, A. Kuznetsov, Y. Gorbenko, A. Pushkar'Ov, Y. Kotukh, K. Kuznetsova, Random S-boxes generation methods for symmetric cryptography, in: 2019 IEEE 2nd Ukraine Conference on Electrical and Computer Engineering, UKRCON 2019 - Proceedings, 2019: pp. 947–950. https://doi.org/10.1109/UKRCON.2019.8879962.

[44] A. Kuznetsov, N. Poluyanenko, E. Frontoni, S. Kandiy, O. Pieshkova, Optimized simulated annealing for efficient generation of highly nonlinear S-boxes, Soft Comput (2023). https://doi.org/10.1007/s00500-023-09334-y.

[45] A. Kuznetsov, E. Frontoni, L. Romeo, N. Poluyanenko, S. Kandiy, K. Kuznetsova, E. Beňová, Optimizing Hill Climbing Algorithm for S-Boxes Generation, Electronics 12 (2023) 2338. https://doi.org/10.3390/electronics12102338.

[46] A. Freyre-Echevarría, I. Martínez-Díaz, C.M.L. Pérez, G. Sosa-Gómez, O. Rojas, Evolving Nonlinear S-Boxes With Improved Theoretical Resilience to Power Attacks, IEEE Access 8 (2020) 202728–202737. https://doi.org/10.1109/ACCESS.2020.3035163.

[47] E.C. Laskari, G.C. Meletiou, M.N. Vrahatis, Utilizing Evolutionary Computation Methods for the Design of S-Boxes, in: 2006 International Conference on Computational Intelligence and Security, 2006: pp. 1299–1302. https://doi.org/10.1109/ICCIAS.2006.295267.

[48] A. Ghosh, S. Das, B. Saha, Chapter 6 - Nature-inspired optimization algorithms, in: A. Ghosh, S. Das, B. Saha (Eds.), Artificial Intelligence in Textile Engineering, Woodhead Publishing, 2024: pp. 171–231. https://doi.org/10.1016/B978-0-443-15395-2.00002-8.

[49] C.-W. Tsai, M.-C. Chiang, Chapter Seven - Genetic algorithm, in: C.-W. Tsai, M.-C. Chiang (Eds.), Handbook of Metaheuristic Algorithms, Academic Press, 2023: pp. 111–138. https://doi.org/10.1016/B978-0-44-319108-4.00020-4.

[50] T. Kapuściński, R.K. Nowicki, C. Napoli, Application of Genetic Algorithms in the Construction of Invertible Substitution Boxes, in: L. Rutkowski, M. Korytkowski, R. Scherer, R. Tadeusiewicz, L.A. Zadeh, J.M. Zurada (Eds.), Artificial Intelligence and Soft Computing, Springer International Publishing, Cham, 2016: pp. 380–391.

https://doi.org/10.1007/978-3-319-39378-0_33.

[51] C.-W. Tsai, M.-C. Chiang, Chapter Sixteen - Local search algorithm, in: C.-W. Tsai, M.-C. Chiang (Eds.), Handbook of Metaheuristic Algorithms, Academic Press, 2023: pp. 351–374. https://doi.org/10.1016/B978-0-44-319108-4.00030-7.

[52] A. Kuznetsov, N. Poluyanenko, S. Kandii, Y. Zaichenko, D. Prokopovich-Tkachenko, T. Katkova, WHS Cost Function for Generating S-boxes, in: 2021 IEEE 8th International Conference on Problems of Infocommunications, Science and Technology, PIC S and T 2021 - Proceedings, 2021: pp. 434–438. https://doi.org/10.1109/PICST54195.2021.9772133.

[53] A. Kuznetsov, O. Potii, N. Poluyanenko, S. Ihnatenko, I. Stelnyk, D. Mialkovsky, Opportunities to minimize hardware and software costs for implementing boolean functions in stream ciphers, International Journal of Computing 18 (2019) 443–452.

[54] A. Kuznetsov, L. Wieclaw, N. Poluyanenko, L. Hamera, S. Kandiy, Y. Lohachova, Optimization of a Simulated Annealing Algorithm for S-Boxes Generating, Sensors 22 (2022) 6073. https://doi.org/10.3390/s22166073.

[55] A. Kuznetsov, M. Karpinski, R. Ziubina, S. Kandiy, E. Frontoni, O. Peliukh, O. Veselska, R. Kozak, Generation of Nonlinear Substitutions by Simulated Annealing Algorithm, Information 14 (2023) 259. https://doi.org/10.3390/info14050259.